\documentclass[12pt,oneside]{article}

\usepackage{amsfonts,amssymb,graphicx}
\usepackage{amsmath}

\usepackage{mdwlist}
\usepackage{enumitem}

\setlist{nolistsep}

\usepackage[bottom]{footmisc}

\usepackage{caption, subfig}

\usepackage{booktabs}

\def\P{ {\rm P} }
\def\E{ {\rm E} }

\def\ra{\rightarrow}

\newcommand{\ulimit}[1]{\underset{#1}{\longrightarrow}}

\def\no{\noindent}
\def\be{\begin{equation}}
\def\ee{\end{equation}}
\def\bea{\begin{eqnarray}}
\def\eea{\end{eqnarray}}

\def\<{\langle}
\def\<{\langle}
\def\>{\rangle}

\def\ds{\displaystyle}

\def\~{\tilde}

\def\a{\alpha}

\def\ds{\displaystyle}

\makeatletter
	\def\blfootnote{\xdef\@thefnmark{}\@footnotetext}
\makeatother

\renewcommand\l{\lambda}

\begin{document}
%

%\hsize=16.0truecm\vsize=22.5truecm\vglue6.3truecm

\begin{center}
%\vspace{1truecm}

{\bf\sc\Large population genetics of gene function \\}

%existence and solution }\\
%\vspace{1cm}
\vskip .5truecm
{Ignacio Gallo \\
%\vspace{.5cm}
%{\small ***institution*** \\ {e-mail: {\em ignacio@cantab.net} } } 
%\today  \\
} 
\end{center}
%\vskip .5 truecm

\blfootnote{email: {\it ignacio@cantab.net}}

%\begin{center}
%{
%{\bf\sc \Large a simple model leading to a relation between ecological and genetic quantities\\
%}
%\vspace{.5truecm}
%\large
%Ignacio Gallo % \\ \today
%}
%\end{center}

%\tableofcontents
%\newpage

%\vskip .3 in

\vskip 1 truecm

\begin{abstract}\noindent
\footnotesize This paper shows that differentiating the lifetimes of two phenotypes independently
		     from their fertility can lead to a qualitative change in the equilibrium of a population: since 
		     survival and reproduction are distinct functional aspects of an organism, this observation contributes 
		     to extend the population-genetical characterisation of biological function. 
		     To support this statement a mathematical relation is derived to link the lifetime ratio $T_1 / T_2$, which
		     parametrizes the different survival ability of two phenotypes, with population variables that quantify the amount
		     of neutral variation underlying a population's phenotypic distribution. 
		      \end{abstract}

\vskip 1 truecm

%
%

%
%		OPENING REMARKS
%

%
%

\noindent During the last decade experimental research has begun using population-genetical 
principles to link the function of genes to their population distribution \cite{nielsen, williamson, sawyer}. 
This closely parallels the way in which statistical thermodynamics
links the macroscopic state of a physical body to the activity of the molecules that form it, and is 
likely to bear similarly significant consequences.

Population genetics has studied the mathematical relation of evolutionary and demographic forces
to the population distribution of alleles for more than a hundred years, and bioinformatics has long used 
conserved sites in comparative statistical data to infer function \cite{nielsen}. More recent works can therefore 
be seen as a joint maturation of these long-standing research threads, which enables to pose 
questions about the relation of gene function to gene distribution
in a more detailed manner than previously possible: the analogous maturation which gave birth 
to statistical thermodynamics consequently allowed to infer the size of a molecule from observable dynamics.

One important point that has been stressed in recent works is that demographic forces can mimic the effect of selection
during evolution, so that it is necessary to account for demography when inferring selection from population statistics. In 
\cite{williamson} the authors consider the effect of a changing population size, 
and show that a population's pattern of variation may be used to infer this type of demographic information, as well as 
to locate functionally important genetic sequences.

The present paper considers a further possibility granted by this approach, which lies in using the different effects
of demography and reproductive selection on gene distribution in order to infer more detailed information about gene function 
itself.

It needs to be stressed that, for the sake of exposition, this paper uses the term ``gene function" as relating
to a univocally defined concept, while yet acknowledging that this concept clearly doesn't exist as such. One 
good way to justify such simplistic use lies in the analogy with the term ``the meaning of a word", which similarly can
sometimes be put to good use, while suffering from an analogous ambiguity of definition.

Life-expectancy, a demographic parameter, is related to function: it reflects, statistically, the 
ability of an organism to perform the tasks which are required in order to survive for certain amount of 
time. If life-expectancy is systematically different for two phenotypes in a given environment, it is 
natural to conclude that this difference is due to the different way in which the two phenotypes 
function in such environment.

Therefore, if we consider survival and reproduction to be two macro-functions performed by 
any living organism, detecting a difference in phenotypic life-expectancy -separately from 
a difference in fertility- must allow to get a two-dimensional description of the gene function
which underlies the phenotypic change. This suggests that including the effect of demographic forces 
may not only be used to infer information about a species' demographic history, but also to get more 
detailed information regarding the specific function that a gene is playing in a given environment, 
using only information which characterises the population as a whole.

This paper shows that life-expectancy can affect the nature of a population's phenotypic
distribution in a way which is qualitatively distinct from the effect of fertility. It is
also shown that if a population contains two phenotypes characterised by different life-expectancies $T_1$
and $T_2$, then the ratio 
$$
\l=\frac{T_1}{T_2}
$$
can be quantitatively estimated through the phenotypes' respective amounts of neutral variation, independently
of all other parameters.

Since a population's dynamics is typically dominated by differences in fertility, standard models tend
to combine the effect a genetic change has on an organism's survival ability with its effect on 
fertility \cite{crowkimura, moran2}. 
In this paper we study the effect of a difference in survival explicitly: we find that this effect is 
indeed small, but that it leads to qualitative changes in a population's equilibrium regime, which are 
likely to generate statistically observable signals at a population as well as at a comparative level.

The paper is organised as follows: in the next section we describe the empirical justification for the structure
of the chosen model, and we then present the model in section 2. Our modelling framework consists of two levels: one
level studies the equilibrium reached by two available phenotypes, and is analysed in section 3. It is at this level 
that we observe a qualitatively novel equilibrium regime that results from differentiating the phenotype lifetimes
separately from their reproductive rates.

Section 4 focuses on the second level of description, which corresponds to the amount of neutral variation 
characterising each of the two considered phenotypes. We are interested in this variation because it allows 
to express the parameter $\l = T_1/T_2$ in terms of statistically observable data.

We conclude the paper by considering the practical limitations of the derived results and briefly outlining possible
further developments.

\section{Empirical background}

%At the phenotypic level we propose a modification on the typical setting of Wright and Moran,
%which consists in a clear variation on the classical stochastic process which we try to motivate
%as satisfactorily as possible. This modification produces

We want to construct a stochastic model for the dynamics of a population consisting of two phenotypes
which differ both in their average amount of offspring (which we call $W_1$ and $W_2$, respectively),
and in their average lifetimes ($T_1$ and $T_2$). To this end, in section~2 we modify the haploid Moran 
model by a qualitative change in its process, which we attempt to justify on intuitive grounds, and 
which we parametrize in terms of a novel parameter, $\ds \l =  T_1/T_2$. 

\begin{figure}[t]
	\centering
	\includegraphics[width=15 cm]{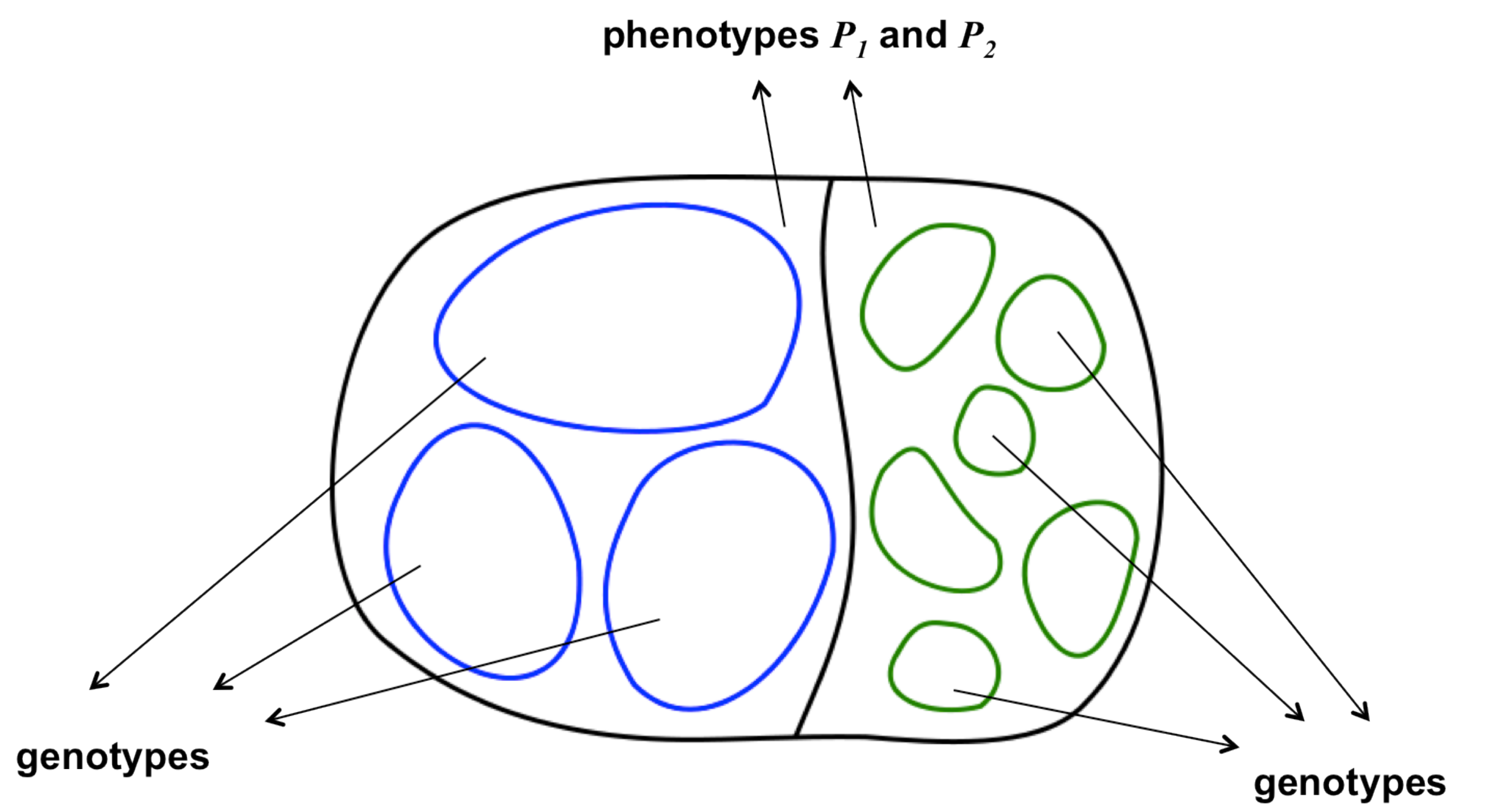}
	\caption{\it Structure of the considered population: organisms carry one of two phenotypes, $P_1$ and $P_2$, each
			of which is coded by many synonymous genotypes that are assumed to be neutral with respect to each other.}
	\label{phenogeno}
\end{figure}

Since the model includes one more parameter than the standard setting, it is desirable to correspondingly 
expand the number of independent quantities that we expect to observe, so to allow
the discrimination among possible causes of specific states of the population. We address this need by considering 
the amount of synonymous variation included in each of our two phenotypes, which, following \cite{williamson}, we consider to be
neutral: we therefore have a population consisting of 
two phenotypes coded by a larger number of genotypes, which are neutral with respect to each other
as long as they give rise to the same phenotype (Figure \ref{phenogeno}).

%\begin{figure}[!ht]

This population structure parallels the structure of variation encountered by Martin Kreitman when studying the
alcohol dehydrogenase locus in {\it D. melanogaster}, where he found that only one of the 43 polymorphic sites 
observed in his sample led to a change in the protein coded by the gene, thus revealing that the gene consists
of only two molecular phenotypes coded by a larger number of genotypes \cite{kreitman}.

Though the model we consider is clearly much too simplistic to apply to a natural population of {\it Drosophila}, Kreitman's 
observation gives empirical justification for building a model which allows only one of the sites of a long genetic 
sequence to lead to a change in phenotype, which would a priori seem to be a very strong assumption. 

We will see that by using such simplification -which is further supported by more extensive studies \cite{berrykreitman}- 
our model allows both to derive a clear characterisation of the phenotypic equilibria (section 3), and to estimate 
the model's novel parameter $\l$ (section 4) in terms of statistically observable quantities, by using variations of 
standard results.

\section{Modelling approach and its relation to the standard setting}

The structure of our model reflects the empirical situation described in the last section: we consider
a population of $N$ organisms carrying genotypes of length $L$, where each genotype-site can take two possible states: 
however, only one of these sites corresponds to a change of phenotype, whereas mutations in other sites
are neutral. It is worth pointing out that this is the same structure of the {\it Drosophila} haplotype data included in the appendix 
of \cite{berrykreitman}, where only mutations in one specific molecular-marker site lead to an amino acid change.

The model therefore includes two levels of description: a phenotypic one, which describes the dynamical
change in the number of individuals carrying the two available phenotypes, and a genotypic level that 
characterises the amount of neutral variation available for each of the two phenotypes.

We use the symbols $P_1$ and $P_2$ to denote our two phenotypes, and we keep the population size fixed at a 
value $N$: we can therefore specify the phenotypic state of the population by a random variable 
$X$ which gives
\begin{eqnarray*}
	X      & : & \textrm{number of individuals with phenotype $P_1$},
\\	
	N - X  & : & \textrm{number of individuals with phenotype $P_2$}.
\end{eqnarray*}

We are interested in studying the stochastic equilibrium reached by a process of death, reproduction 
and reversible mutation, where mutation happens with the same probability $u$ in both directions, and for all the 
available sites, including the phenotypically-linked one.

The assumption of a mutation rate which is both symmetric and site-independent is very idealised, and even more
importantly, an equilibrium due to reversible mutation is not generally considered to be relevant for generic 
mutations \cite{hartlclark}. 

Due to its simplicity, however, the chosen setting allows to show with remarkable clarity 
that an explicit difference in the phenotype lifetimes leads to a qualitative novel type
of equilibrium state: this is the aim of this paper, since it is in this novelty that we see 
potential to extend the standard population-genetical
characterisation of biological function; having a full characterisation of this elementary case should be useful when 
considering more realistic and analytically challenging situations.

In the next subsection we therefore describe a variation of the Moran process which, as we will attempt to 
justify on intuitive grounds, provides a good model for phenotypes that are allowed to differ independently in lifetime 
and average number of offspring. It is interesting to point out that Moran introduced in Ref. \cite{moran2} a variation of his 
original model that implements reproductive selection by differentiating his alleles' lifetimes while keeping the instantaneous 
reproductive rates the same for all his phenotypes (thus differentiating their life-long reproductive yields): Moran remarks that considering 
lifetime and reproductive differences separately would {\it almost certainly} make 
no difference for the equilibrium distribution. Our implementation of the phenotypic difference comes as a natural 
extension of Moran's, and our claim is that, though the subtlety of this extension's effect roughly confirms his intuition, 
the qualitative nature of this change provides considerable descriptive potential.

As stressed before, when introducing a quantity $\l$ to parametrize the differentiation in the lifetimes, it becomes 
desirable to extend the set of quantities that we expect to observe in the statistics of the population: in other words, though the quantity
$$
x = \frac{X}{N}
$$
fully characterises our population's phenotypic state, we can gather further information by looking at how the $X$ individuals
carrying phenotype $P_1$ are partitioned into synonymous genotypes, and similarly for the $N-X$ individuals carrying 
phenotype $P_2$.

A natural intuitive choice to characterise neutral variation in the two phenotypes would be to count the actual number of 
synonymous genotypes present for each: however, as remarked in \cite{crowkimura}, an interesting alternative is to use the inbreeding 
coefficient concept.

The inbreeding coefficient is typically used for diploid organisms, since it is defined as the probability that a given genetic locus is homozygous:
that is, it is the probability that for a diploid organism chosen at random from a population, the two alleles that the organism
contains at such locus are found to be identical. However, under the assumption of random mating, this turns out to be equivalent
to the probability that any two alleles drawn at random from the population are identical, regardless of the separation into organisms.

The latter definition makes the quantity relevant to haploid populations, and it turns out to be a more analytically
accessible one than the aforementioned {\em actual number} of synonymous genotypes, at least for the estimation purpose considered in section \ref{estimation},
in which we generalise a result taken from \cite{kimuracrow}. The inbreeding coefficient also provides an effective approximation
to the actual number of synonymous alleles, and it has been suggested to be more empirically accessible than the 
latter quantity \cite{crowkimura}.

\subsection{Phenotypic level}

Here we describe the process through which we model the change in
our population's phenotypes, and make our attempt to justify the modelling choice through
intuition: the observable outcome of this process is analysed in section \ref{equilibrium}.

Studies using population genetics to infer function are typically based on the Wright model \cite{nielsen, williamson}, 
which describes the stochastic change of a population at discrete non-overlapping generations, 
and it is customary to describe the allele dynamics by using a continuous approximation to this process.
There is, however, a somewhat paradoxical aspect to this standard modelling approach: whereas the Wright model 
describes the population as changing in discrete generations which might last for a considerable amount of time, the 
continuous approximation requires such generations to be taken of vanishing duration.

This assumption is well justified by the fact that processes are often considered 
to be taking place on an evolutionary time scale, which is much longer than the generation time, 
as well as by the fact that if one assumes organisms not to be subject to ageing, which 
is virtually unavoidable at the simplest level of description, the Wright model is formally equivalent 
to a process of death and birth \cite{cannings}.

\captionsetup[subfloat]{nearskip=15pt,captionskip=4pt}

%\begin{figure}[!ht]
\begin{figure}%[t]
	\centering
	\subfloat[][]{\includegraphics[width=12 cm]{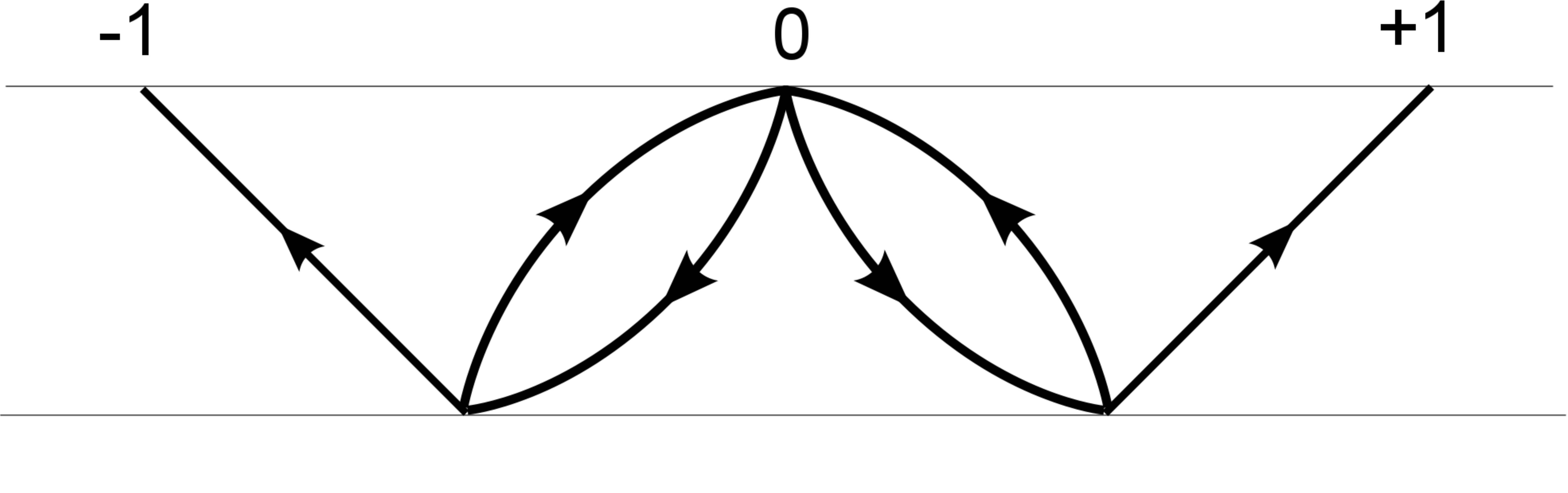}}%
	\qquad
%	\subfloat[][]{\includegraphics[width=12 cm]{pictures/presentModel.jpg}}
	\subfloat[][]{\includegraphics[width=12 cm]{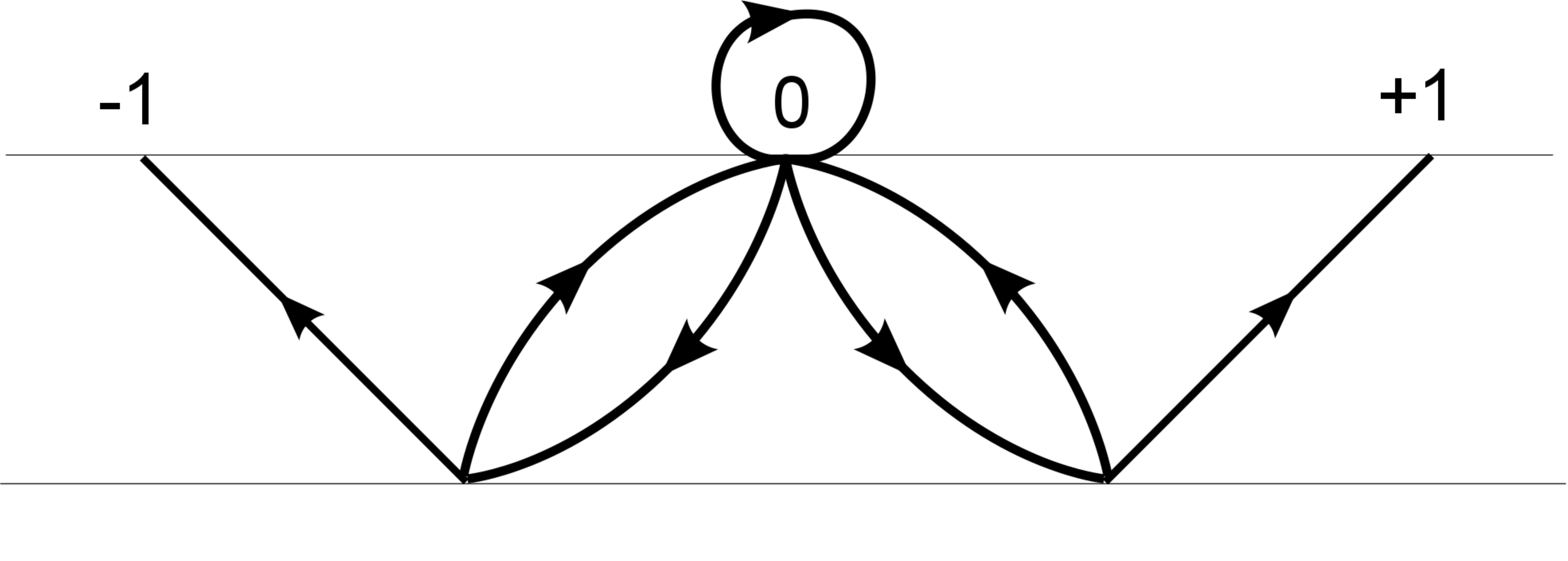}}
	\caption{\it Here we illustrate the difference between the (a) Moran process and (b) the process considered in this paper.
		     At every time interval in (a) an organism is chosen to die, and a new one is chosen to replace it: 
		     the types of the dead and newborn organisms determine change in phenotype $P_1$'s frequency $X$ as $-1$, $0$ or $1$.
		     The loop in (b) shows that in the present model time flows according to an external reference: this allows to 
		     characterise the different nature of birth and death events.}%
	\label{MoranAndMe}%
\end{figure}

On the other hand, the point of view of this paper is that, though suitable given the typical assumptions,
the reliance on a discrete generation setting might hinder the consideration of
relevant modifications: here we propose a modification to the Moran model that stems from features of
an instantaneous event, which are prohibitively  difficult to visualise at a generation level.

The Moran model describes the process of change in a population consisting of two types 
of organism.
A time interval in this model is defined by the occurrence of a death event, followed by a birth event. 
It is sensible, in principle, to define time through these events, since these change the state 
of the population, which is the object under study.

There is a reason, however, to regard time as flowing according to an external frame of reference,
 thus including time intervals during which no population events happen:
death events may be thought to take place at a rate determined
by the physiology of the organisms; due to competition, however, birth events may be thought to happen
instantaneously when the death of an individual makes a safe spot available in the environment. 

In fact, both the Wright and the Moran model set the population size to be fixed at a value $N$, which represents 
the environment's carrying capacity: the meaning of this constraint is that a death event should 
be interpreted as one which vacates an environmental safe spot, and we argue that the presence of competition
makes it intuitively admissible to model the subsequent birth event as happening instantaneously as soon as the environmental
spot becomes available.

As a consequence, having an external time reference allows to describe more adequately the interplay 
between the two different types of competition which characterise 1) an organism's struggle to survive, and thus 
to preserve its environmental spot, and 2) the reproductive struggle to occupy all spots as soon as they become available.

According to this interpretation, all organisms in a population might be thought to be playing a waiting game similar 
to the children's game ``musical chairs" where $N-1$ chairs are available for $N$ children to sit on when the music 
stops. This process has already been considered in an evolutionary setting in \cite{binmore}: in our case, rather
than focusing on modelling the competitive game which determines the allocation of an available spot, we consider
this allocation to happen trivially and instantaneously, and we focus on the waiting game itself.

In practice, our modification of the Moran model consists of a process which allows {\it at most} one death-birth event 
per time interval rather than {\it exactly} one as in the original model  \cite{moran}: the two diagrams in Figure \ref{MoranAndMe} 
illustrate this difference. 

Figure \ref{MoranAndMe}(a) shows the change in the phenotype frequency $X$ during a time 
interval in the original Moran model: an individual is chosen at random from the population and killed, and then replaced by a 
new individual. The change in $X$ is then determined by the phenotypic identity of the dead and of the newborn.

Differently from Figure \ref{MoranAndMe}(a), Figure \ref{MoranAndMe}(b) (which describes our process) contains a loop at the origin of the diagram.
This formalises the different nature of the death and birth events: at a given
time instant no organisms might die; when a death does happen, however, a birth systematically follows instantaneously.

This modelling choice is arbitrary: contrarily to our assumption, following a death competitive conflicts between organisms 
might lead to a substantial delay in the allocation of the newly-vacated spot, and this could considerably change the nature of the 
process. This objection, however, only highlights the descriptive potential of a modelling approach that uses intuition to consider 
fundamental population events in some detail, 
an approach for which a considerable gap exists in the mathematical biology literature: here we 
look at the consequences of a simple such possibility.

%As we will see
%this difference has both a qualitative effect on the population's phenotypical equilibrium state, and leads to quantitative
%consequences on each phenotype's neutral variation that can be used to infer the value of the ratio between the
%two phenotypes' average lifetimes from statistical observable data, thus providing relatively detailed functional
%information on phenotypes from population level observations.

\begin{figure}[t]
	\centering
	\includegraphics[width=12 cm]{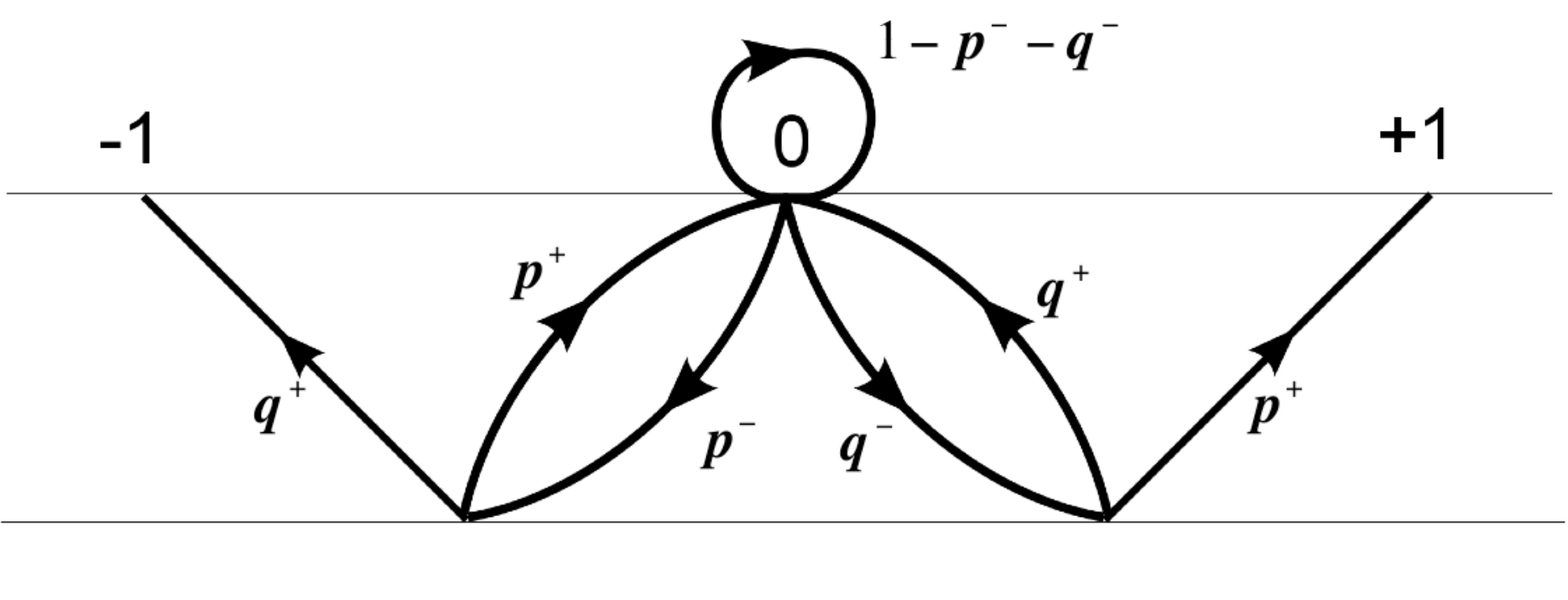}
	\caption{\it Transition probabilities for the fundamental population events in our process: $p^-$ and $p^+$ correspond
		     to death and birth (after mutation) of organisms with phenotype $P_1$. Similarly we have $q^-$ and $q^+$
		     for $P_2$, and the existence of the loop at the origin of the diagram is due to that in general $p^- + q^- < 1$.  }
	\label{transprob}
\end{figure}

The model is therefore defined by the transition probabilities $p^-$, $q^-$, $p^+$ and $q^+$  in Figure \ref{transprob},
which are in turn derived from the life-cycles of the two organisms.

Transition probability $p^-$ corresponds to the event that an organism with phenotype $P_1$ dies, whereas $q^-$ 
corresponds to the same event for phenotype $P_2$.

We denote the relative frequency of phenotype $P_1$ by $x=X/N$, and its average lifetime by $T_1$: under the assumption 
that each organisms is reproductively mature at birth and is not subject to ageing, we have that
\begin{equation}\label{deathprobs}
	p^- = \frac{x}{T_1}, \quad \textrm{and} \quad q^- = \frac{1-x}{T_2}.
\end{equation}

 In this paper we refrain from giving a fully detailed derivation of these formulas: the issues involved
in the rigorous foundation of this level of modelling are problematic, and this is indeed related to the fact that variations 
such as (\ref{deathprobs}) are not often encountered in the literature. This paper rather tackles foundational issues by
proposing (\ref{deathprobs}) as a specific variation of the standard approach.

The technical aspects of the derivation of (\ref{deathprobs}) are not, however, fundamentally different from those encountered in
the Wright and the Moran models, and the connection can be intuitively clarified by the following observation. The quantity $p^{\,-} \,$
is the product of {\em 1)} the Moran-like probability that an organism carrying phenotype $P_1$ is chosen to die $\ds (x=X/N)$, and 
{\em 2)} the probability that it {\em actually} dies. The latter probability, which we can call $\delta_1$, corresponds to the fact that in our
model organisms are always given a chance to survive: this can be given a more fundamental justification if one considers a model
in which an arbitrary number of organisms can die at any given time interval. We shall, however, refrain from pursuing this 
line of reasoning further, and leave it for a more specific future work.

We want our model's parameters to correspond to biological features: assuming that our organisms are not subject to 
ageing, or to environmental fluctuations, we have that their average lifespan is equal to the mean of a geometric distribution
with parameter $\delta_1$ (for phenotype $P_1$), which leads to
$$
T_1 = \frac{1}{\delta_1}.
$$
This gives $p^-$ in (\ref{deathprobs}), and the same reasoning applies to $q^-$.

In view of (\ref{deathprobs}) we have that in general
$$
p^- + q^- < 1,
$$
and this is the cause of the qualitative effect arising from differentiating the phenotypic lifespans, which gives the model's novelty.

The second biological feature which we assign to our phenotypes is the average number of offspring produced by an organism
during its entire lifetime, which we denote by $W_1$ and $W_2$ for phenotypes $P_1$ and $P_2$, respectively.

Transition probabilities $p^+$ and $q^+$ are obtained by considering elementary events in a similar way as for $p^-$ and $q^-$,
taking into account that reproduction involves also mutation, which we model as happening with probability $u$ in both directions.

Under these life-cycle conditions it can be shown that
$$
p^+ = \frac{ \ds \frac{W_1}{T_1} \, (1-u) \, x + \frac{W_2}{T_2} \, u \, (1-x) }{ \ds \frac{W_1}{T_1} \,  x + \frac{W_2}{T_2} (1-x) },
$$
and
$$
q^+ = \frac{ \ds \frac{W_1}{T_1} \, u \, x + \frac{W_2}{T_2} (1-u) (1-x)    }{ \ds \frac{W_1}{T_1} \, x + \frac{W_2}{T_2} (1-x) }.
$$

The reason for the denominators in $p^+$ and $q^+$ is that, 
as we stressed before, a reproduction event is assumed to happen instantaneously when a death event vacates an environmental spot,
so that a ``death followed by no birth" is not considered to be a possible event. This determines the ``musical-chairs" nature of the model, 
which we claim to be a particularly insightful way of modelling a process of competition \cite{binmore}, and 
which quantitatively corresponds to 
$$
p^+ + q^+ =1.
$$

%Though the justification to this variant is arguably meaningful, the , as in any model, 

%Granted this arbitrariness, we will now turn to consider the consequence

%We expect the genetic sequences related to phenotypes whose distribution does not become trivial in the large population
%size to be the most informative. 

We are particularly interested in including parameter values for which the equilibrium distribution does not become trivial in the limit of large 
population size, and to this end we employ the following asymptotic scalings for the mutation parameter
$$
N \, u \ulimit{N \ra \infty} \theta,
$$
and for reproductive selection parameter
$$
N \, \Bigg ( \frac{W_1}{W_2} -1 \Bigg ) \ulimit{N \ra \infty} s,
$$
which we use as definitions for the rescaled parameters $\theta$ and $s$. For convenience we also use the parameter $\l$ 
for the ratio between lifetimes:
$$
\l = \frac{T_1}{T_2}.
$$

In section \ref{equilibrium} we will need the first two moments of the change in the variable $X$ at a given time to
write down the large population size limit for the equilibrium distribution attained by the phenotypes. To this end, after defining 
$$
\Delta X (x) = X_{t+1} - X_t \, ,
$$
we need to compute quantities $M(x)  = \E \, \big [ \Delta X (x) \big ] $ and $V(x) = \E \, \Big [ ( \Delta X (x) \big )^2  \Big ] $ in
terms of the transition probabilities defined above. The functional dependence on $x$ shows that this moments 
are computed conditionally on the relative frequency of phenotype $P_1$ being equal to $x=X/N$: for convenience, however, 
from now on we drop the $x$ dependence from the notation.

Using inspection on Figure \ref{transprob} we find that
$$
M = \lim_{N \ra \infty } \;  ( q^- p^+ - p^- q^+ ),
$$
and that
$$
V = \lim_{N \ra \infty } \;  ( q^- p^+ + p^- q^+ ),
$$
which in terms of the asymptotic parameters gives
$$
M = \frac{1}{N T_1} \cdot \frac{ \theta \, \l^2 \, (1-x)^2 + \l \, s \, x (1-x) - \theta \, x^2 }{ x + \l (1-x) },
$$
and
$$
V = \frac{ 2 \l }{ N T_1 } \cdot \frac{ x \, (1-x) }{ x + \l (1-x) }.
$$

We see that the novelty of the model is nicely shown algebraically by the presence of a factor $(x + \l (1-x))$
in the denominator of both $M$ and $V$, its presence in the latter being particularly significant for the form
of the equilibrium distribution: we discuss the analytic consequences of this in section \ref{equilibrium}.

%Figure \ref{process} shows how the random step appears within the whole stochastic process.
%
%\begin{figure}[!ht]
%	\centering
%	\includegraphics[width=15 cm]{pictures/process.eps}%
%	\caption{blac bla bla}
%	\label{process}
%\end{figure}

\subsection{Neutral variation}\label{neutral}

Underlying the process of change in the phenotypic frequencies we have the process of creation
of new neutral mutations to phenotypes $P_1$ and $P_2$, and of their stochastic loss.

As mentioned in the introduction to this section, we model each genotype as a sequence
of $L$ two-state sites, which includes a site (the ``phenotypically-linked" site) whose mutation causes the change between 
phenotypes $P_1$ and $P_2$, and
for the sake of simplicity we make the rather strong assumption that mutation happens with
same probability $u$ at all sites, and in both directions: 
therefore, the probability of mutation $u$ relevant to the phenotypic equilibrium also parametrizes the
amount of neutral variation for the two phenotypes.

%We furthermore assume that each mutation in a neutral site of a newborn
%genotype leads to the creation of a new neutral mutation for the parent phenotype, 

%We therefore have that

Like we said before, rather than using the actual number of neutral genotypes into which each phenotype
is partitioned, we choose to characterise neutral variation by the inbreeding coefficient. For
a population of haploids, such as the one we consider, the inbreeding coefficient can be defined as the 
probability that two genotypes drawn at random from the population are identical.

Therefore, in addition to random variable $x$ that characterises the population's phenotypic
distribution, we define the two quantities
\begin{eqnarray*}
	F_1 = \textrm{probability that two organisms with phenotype $P_1$ have the same genotype},
\\
	F_2 = \textrm{probability that two organisms with phenotype $P_2$ have the same genotype}.
\end{eqnarray*}

In section \ref{estimation} we find an explicit formula for the new parameter $\l$ in terms of 
combined moments of quantities $F_1$, $F_2$ and $x$: in this paper's point of view such 
a relation contributes to extend the population-genetical characterisation of biological function.

%It has been suggested that the inbreeding coefficient may be more empirically accessible
%quantity than the actual number of neutral alleles within a population \cite{crowkimura}.
%Another major reason to consider this quantity is that it is analytically treatable using simple 
%intuitive methods, which suggests its compatibility with the process.

Kimura and Crow \cite{kimuracrow} find that in an ``infinite alleles" model, which compared to the model
presented here may be thought of as one where only one phenotype exists, and where genotypes
have an infinite number of sites, the inbreeding coefficient is on average equal to
$$
\< F \> \approx \frac{1}{ 2 \, N \, u+1}.
$$
In section \ref{estimation} we generalise their calculation to include our case, and see how
the result can be used to express the parameter $ \l = T_1 / T_2 $ in terms of statistically 
observable quantities.

\section{Phenotypic equilibrium}\label{equilibrium}

Here we describe the equilibrium distribution attained by the population's phenotypes $P_1$ and $P_2$
under our process of selection and reversible mutation. 
It is worth stressing immediately that the qualitative novelty of including the differentiation of phenotype
lifetimes manifests itself analytically in the probability density function of $P_1$'s relative frequency $x$,
$$
\phi(x)= C e^{\a \, x} \, x^{\l \, \theta-1}  (1-x)^{\theta/\l-1} \, \big( x + \l (1-x) \big), 
$$
through the factor $( x + \l (1-x) )$, which is not usually seen in population genetics models. When $\l =1$
this factor is equal to one, and we recover the typical equilibrium distribution for a haploid population under
reversible mutation and selection.

\subsection{Form of the equilibrium distribution}

\captionsetup[subfloat]{nearskip=4pt,captionskip=4pt}

\begin{figure}%[t]
	\centering
	\subfloat[][]{\includegraphics[width=6 cm]{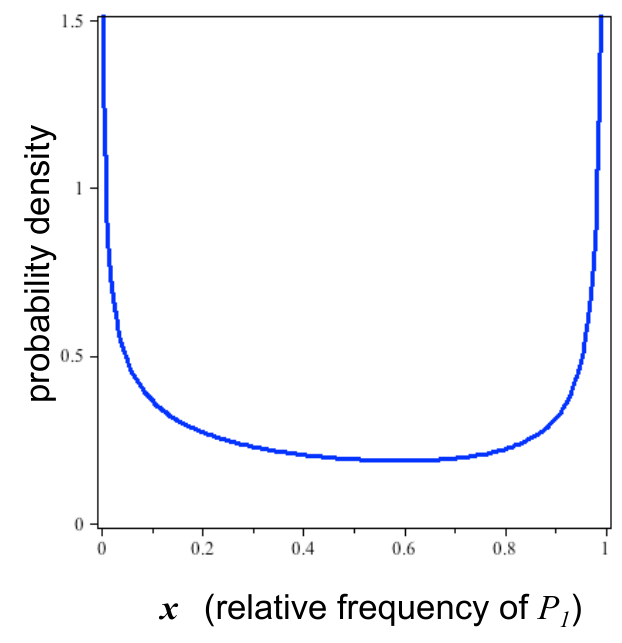}}%
	\qquad
	\subfloat[][]{\includegraphics[width=6.15 cm]{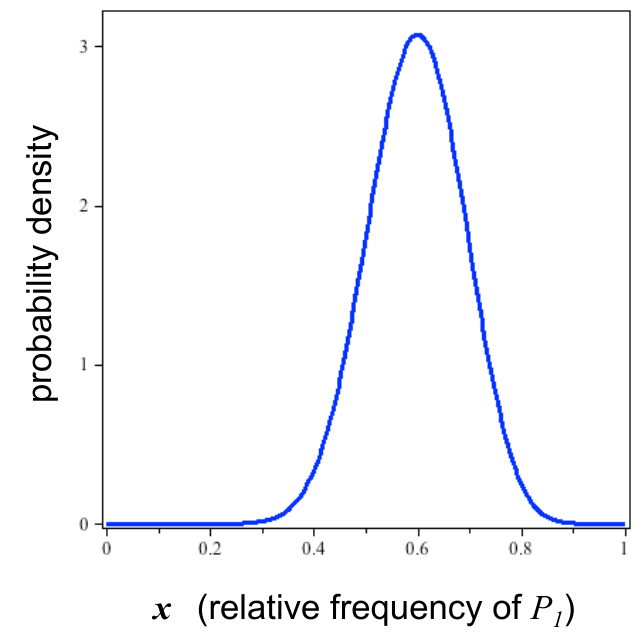}}
	\qquad
	\subfloat[][]{\includegraphics[width=6 cm]{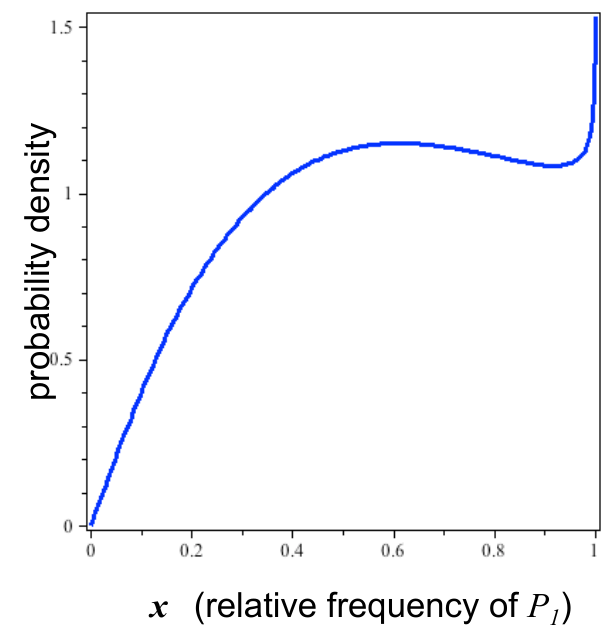}}
	\caption{\it These are the three basic types of equilibrium distribution which can be attained at a phenotypic level, and 
		   which correspond to different intensities of mutation in the following way: (a) $N \, u \ll 1$, (b) $N \, u \gg 1$, (c) $N \, u \approx 1$. 
		   Regime (c) is caused by the differentiation of the phenotype lifetimes, and appears when $\l \neq 1$.}%
	\label{regimes}%
\end{figure}

The form of the phenotypic equilibrium distribution can be obtained in a large population size approximation by 
using Wright's formula
\begin{equation}\label{wright}
	\phi(x)= \frac{C}{V(x)} \, \exp \Bigg \{ 2 \int \frac{M(x)}{V(x)} dx \Bigg \}  ,
\end{equation}
where $C$ is a normalisation constant. 

This formula was derived by Wright \cite{wright} for a process divided into non-overlapping generations, and
it has been proved by Moran to be applicable to various versions of his model \cite{moran2}. More importantly for our case, 
Cannings \cite{cannings} has shown by a concise observation that an overlapping generations model can be considered formally
equivalent to a non-overlapping one as long as organisms are not subject to ageing, and this allows us to use 
approximation (\ref{wright}) in its full generality.

According to the last section our model gives
$$
M = \frac{1}{N T_1} \cdot \frac{ \theta \, \l^2 \, (1-x)^2 + \l \, s \, x (1-x) - \theta x^2 }{ x + \l (1-x) },
$$
and
$$
V = \frac{ 2 \l }{ N T_1 } \cdot \frac{ x (1-x) }{ x + \l (1-x) },
$$
so our model's equilibrium $\phi$ for $P_1$'s relative frequency $x=X/N$ takes the following form
$$
\phi(x)= C e^{\a \, x} \, x^{\l \, \theta-1}  (1-x)^{\theta/\l-1} \, \big( x + \l (1-x) \big), 
$$
where
$$
\a = s +  \theta \, \bigg( \frac{1}{\l} - \l   \bigg),
$$
and $C$ is the normalization constant which ensures that
$$
\int_0^1 \phi(x) dx = 1.
$$

We see that for $\l=1$ (that is, when $T_1=T_2$) we recover the equilibrium distribution for a typical 
haploid random drift model with reversible mutation. When $\l \neq 1$ we have that the factor $( x + \l (1-x) )$ 
produces a qualitative difference in the shape of the distribution, though this only happens for values 
of $\theta = N \, u$ close to one: Figure \ref{regimes} shows the shape of the equilibrium distribution 
for different values of the mutation parameter.

We find that the there is an intermediate regime (Figure \ref{regimes}(c)) between the typical low-mutation U-shape (Figure \ref{regimes}(a)) 
and the high-mutation bell-shape (Figure \ref{regimes}(b)). The intermediate regime admits two stationary points, 
and as a consequence becomes bimodal. In general, we have that for any parameter value the type of equilibrium can be
characterised in terms of the number of stationary points which the distribution exhibits: we carry out such characterisation in the
next subsection.

\subsection{Diagram characterising the equilibrium population's modes}

In the last subsection we saw how a difference in the lifetimes of our two phenotypes can lead to
a new type of equilibrium regime for the population, which consists of a hybrid between the classical
low and high mutation regimes: this new regime exhibits both a local probability maximum (as in the 
high-mutation case) and a maximum at the boundary (as in the low-mutation regime).

%We may summarise the possible regimes by drawing a diagram showing the distribution's stationary
%points for different parameter values: this type of characterisation is reminiscent of the phase diagrams 
%familiar to the statistical study of thermodynamic systems, with the important distinction that in our case the object
%on which we are being informed is of a different nature. A thermodynamic ``phase'' is a state of a system
%which, though modelled as random, is typically shown to be asymptotically deterministic for large system sizes.
%
%The equilibrium state of a biological population, even in our idealised approximation, remains stochastic for large 
%population sizes: therefore rather than being informing us about the asymptotically deterministic
%states available, an equilibrium description informs us about the modes of an intrinsically random equilibrium state, 
%[due to the balance between mutation and population size (this type of balance being absent in physical systems)].
%
%One important consequence of this point of view is that, even if the balance among the biological forces at play
%were sufficiently stable to admit our equilibrium description, the population would expected to fluctuate among
%the available modes far more often than how a thermodynamic system is expected to shift about its available phases.
%
%
%Keeping these considerations in mind, it is clearly useful to construct a diagram summarising in a concise 
%way the available qualitative types of equilibrium regime arising at different parameter values. 

Looking at the the shapes of the equilibrium distributions in Figure \ref{regimes}, % (which turn out to be the only possible ones)
we see that these shapes can be well classified in terms of the number and nature of their stationary points, which in turn determine the
number and location of the distribution maxima, or {\it modes}.

Figure \ref{regimes}(a) has only one stationary point, which is a minimum, and this implies the existence of two modes
at $x=0$ and $x=1$ (the probability density in fact diverges to infinity at these boundary points, though this singularity 
does not affect
the possibility of normalising the distribution): this characterises the regime of low mutation as one where the
population polarises about one of the phenotypes, and rarely switches to the other.

Figure \ref{regimes}(b) also has only one stationary point, which is a local maximum: this suggests that
the dynamics of the population in this regime will typically be one where a mixed phenotypic state fluctuates around 
this maximum.

Figure \ref{regimes}(c) shows a novelty of the present model: we have two stationary points, a local maximum and 
a local minimum, which implies the existence of a second mode also at one of the boundaries.
This suggests that the dynamics will tend to polarise the population around one specific phenotype:
the existence of the local maximum, however, suggests that this state of polarisation should be periodically lost in favour
of a mixed configuration for the two phenotypes, and that this switch should happen considerably more 
often than the switch between the polarised states of (a). This, however, can only be elucidated by considering the model's
dynamics explicitly, and could be the topic of a future work.

In order to understand how parameter values relate to cases (a), (b) and (c), we use the probability distribution function 
we obtained in the last section to locate the stationary points of the distribution
$\phi$.

\begin{figure}[t]
	\centering
%	\subfloat[][]{\includegraphics[width=7.1 cm]{phaseClassic2.png}}%
%	\qquad
%	\subfloat[][]{\includegraphics[width=7 cm]{phaseDeform2.png}}
	\subfloat[][]{\includegraphics[width=7.1 cm]{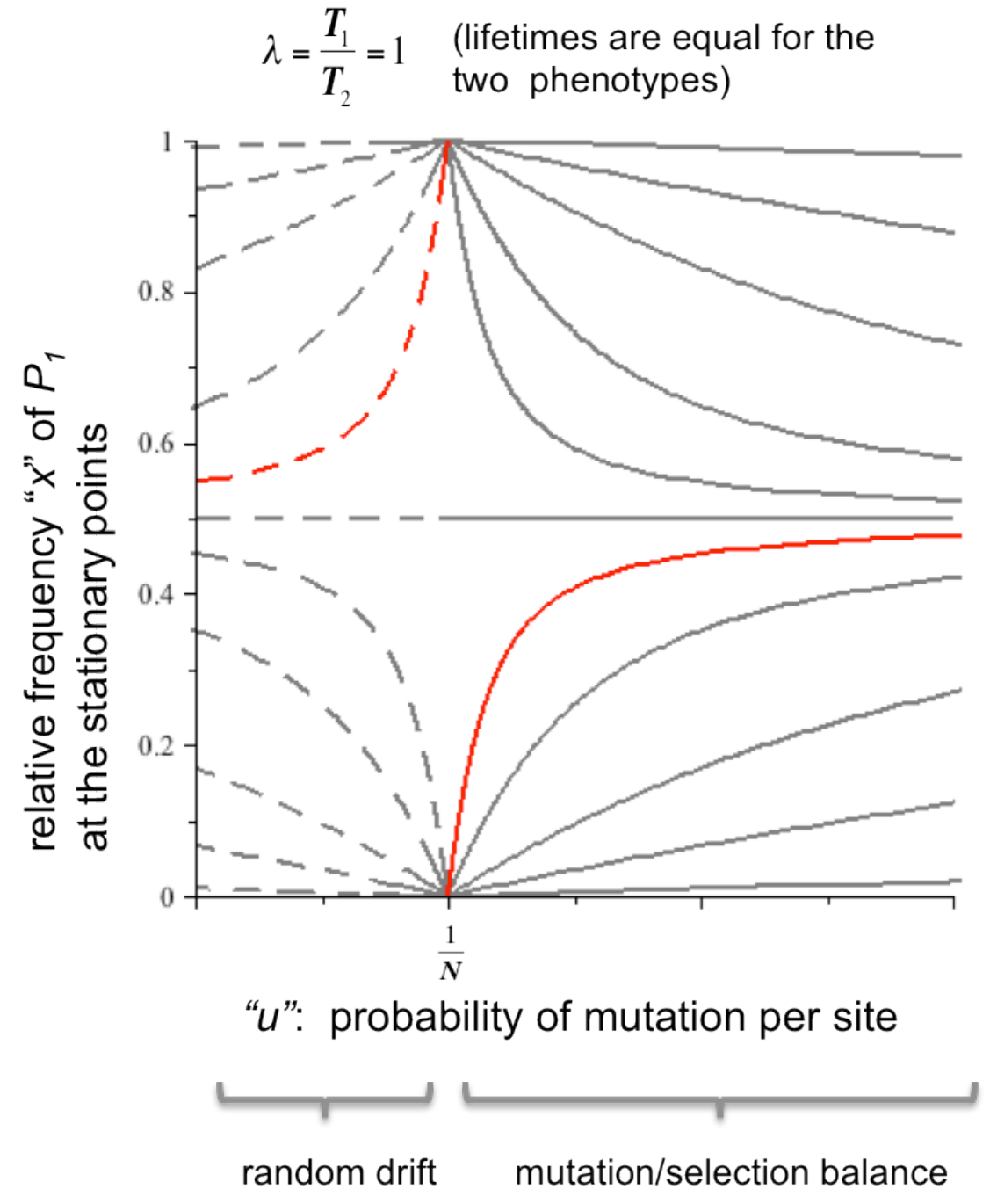}}%
	\qquad
	\subfloat[][]{\includegraphics[width=7 cm]{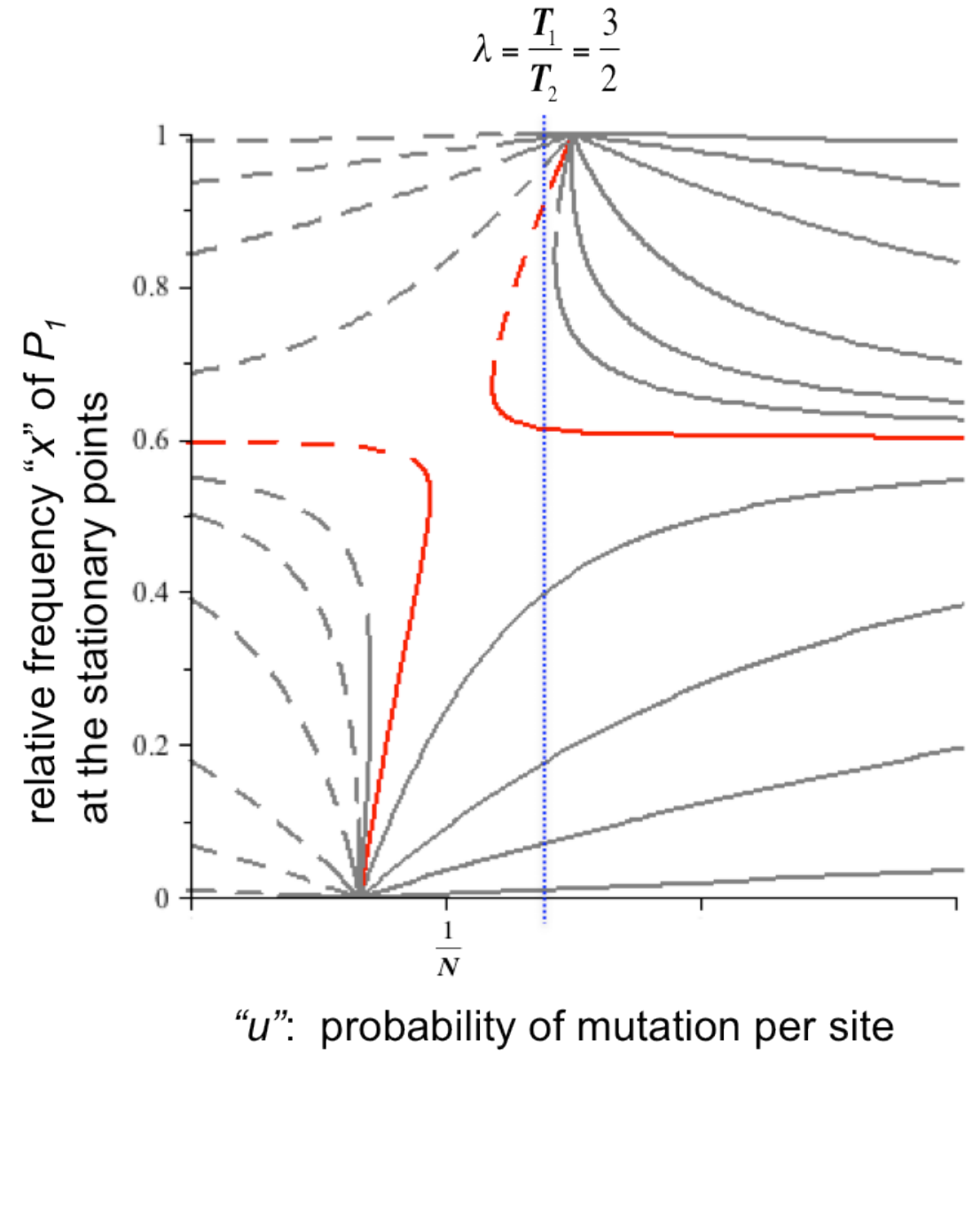}}
	\caption{ \it Stationary points for the equilibrium distribution
			(a) in the classical case where $\l=T_1/T_2 = 1$, and (b) for $\l=T_1/T_2 = 3/2$.
			Each line corresponds to a different value of the selection coefficient ``$\,s$": dashed
			lines are for local minima, solid lines for local maxima.
			When $\l \neq 1$ bimodal equilibrium states exist: the dotted line corresponds to the
			value of the mutation probability ``$\,u$" which gives rise to the distribution in 
			Figure \ref{regimes}(c), for the value of the selection coefficient ``\;$s$" corresponding
			to the line highlighted in red.}%
			
	\label{phase}%
\end{figure}

The condition
$$
\frac{ d \, \phi(x) }{d \, x} = 0,
$$
which gives the stationary points, leads to the following rational equation for $x$:
\begin{equation}\label{modeEqn}
	\bigg ( \a + \frac{a}{ x} - \frac{b}{1-x}  \bigg ) \big ( x + \l \, (1-x) \big ) + 1 - \l = 0,
\end{equation}
where
$$
a = \l \, \theta -1, \quad   b  = \frac{\theta}{\l} - 1,  \quad  \textrm{and}   \quad  \a = s + \theta \, \bigg ( \frac{1}{ \l } - \l \bigg ).
$$
Multiplying equation (\ref{modeEqn}) by factors $x$ and $(1-x)$, we obtain a polynomial
equation of degree 3, which admits three solutions. We are, however, only interested in solutions lying on 
the real interval going from $0$ to $1$, since these correspond to meaningful values for the relative 
frequency $x$. 

Rather than solving the cubic in $x$, we can use (\ref{modeEqn}) to find a functional expression for 
$\theta = N \, u$ (for which equation (\ref{modeEqn}) is linear). By considering $\theta$ 
as a function of $x$, and looking at this function for different values
of parameters $s$ (reproductive selection) and $\l$ (the lifetime ratio), we get a full characterisation 
of the distribution's stationary points and, as a consequence, of its modes: we do this in Figure \ref{phase}.

Figure \ref{phase} shows the stationary points for the equilibrium distribution: the two pictures correspond to
the two different cases $\l=1$ and $\l \neq 1$. Different lines correspond to different values of the selection coefficient $s$,
for which they give the dependence of the position of the stationary points on the mutation probability $u$. 
Dashed lines correspond to local minima and solid lines to local maxima.

Figure \ref{phase}(a) shows the classical
situation where $T_1=T_2$ ($\l=1$). 
As expected, we find an abrupt transition in the nature of the stationary points when the
mutation probability $u=1/N$, at which value the equilibrium distribution turns from being U-shaped to being bell-shaped: 
however, the diagram shows that all parameter values except $u=1/N$
lead to only one stationary point in the equilibrium distribution. Highlighted in red we see the functional dependence
of the unique stationary point on the mutation probability, for a particular value of the reproductive selection coefficient $s$.

%\begin{figure}[t]
%	\centering
%	\subfloat[][]{\includegraphics[width=7 cm]{pictures/phaseClassic.eps}}%
%	\qquad
%	\subfloat[][]{\includegraphics[width=7 cm]{pictures/phaseDeform.eps}}
%	\caption{This picture shows a summary of the equilibrium distribution stationary points
%			a) in the classical case where $T_1/T_2 = 1$, and b) for $T_1/T_2 = 3/2$}%
%	\label{phase}%
%\end{figure}

Figure \ref{phase}(b) shows the same diagram for $\l = T_1 / T_2 = 3/2 $: we see highlighted in red
the dependence of the equilibrium distribution's stationary points on the mutation probability, for the same
value of reproductive selection $s$ used in for the red curve in Figure \ref{phase}(a). 
The diagram shows how near $u=1/N$ there are regions where two stationary points coexist for the same
distribution, a situation which is not encountered in classical population genetics models for haploid 
populations. The dotted line, which shows an example of this, corresponds to the regime in 
Figure \ref{regimes}(c).

An important fact which we learn from Figure \ref{phase} is that the diagram for $\l=1$ is not robust with respect
to changes in the parameter values, whereas
it is robust for any other value of $\l$. This means that for any $\l \neq 1$ the diagram will exhibit regimes
where the equilibrium distribution admits more than one stationary point, and the transition between the 
different qualitative types of equilibrium will come about through the same type of bifurcations which we see
in Figure \ref{phase}(b).

\section{Estimation of $\ds \l= \frac{T_1}{T_2}$}\label{estimation}

In this section we derive the parameter $\l = T_1/T_2$ from population data, and in particular
from statistical quantities characterising the amount of neutral variation underlying phenotypes $P_1$ and $P_2$.
As mentioned above, in order to quantify the amount of neutral mutation we use the inbreeding coefficient
concept, by defining the following two quantities
\begin{eqnarray*}
	F_1 = \textrm{probability that two organisms with phenotype $P_1$ have the same genotype},
\\
	F_2 = \textrm{probability that two organisms with phenotype $P_2$ have the same genotype}.
\end{eqnarray*}

%The inbreeding coefficient is often defined as the probability of a diploid individual being homozygous for a 
%certain allele.
%
%However, under random mating this quantity is equal to the probability of two alleles randomly drawn from the
%population being equal, regardless of the separation into organisms.
%
%This implies that the quantity can be defined meaningfully for a haploid population, and here we argue that it is
%an especially meaningful one.

An important reason for using this approach is that it allows to extend an intuitive result obtained
by Kimura and Crow in \cite{kimuracrow}, where the equilibrium value for the inbreeding coefficient was computed 
for a population consisting of only one phenotype, rather than of two phenotypes like in our case.

The basic observation that allows Kimura and Crow's calculation is that, if we denote the inbreeding coefficient for their
unique phenotype by $F$, and if we assume the population changes according to Wright's process, 
in the absence of mutation the value of $F$ changes according to the following equation:
\begin{equation}\label{basic}
	F(t+1) = \frac{1}{N} + \bigg ( 1 - \frac{1}{N} \bigg ) F(t).
\end{equation}
The intuitive reason for this is that at generation $t+1$ any two individuals have a probability $1/N$ of being born
from the same parent: if they do, in the absence of mutation they share the same genotype with probability $1$, which
gives the first term on the left hand side of (\ref{basic}).

In the presence of mutation, assuming that each mutation generates a mutation which previously didn't exist, 
$$
	F(t+1) = \bigg \{ \frac{1}{N} + \bigg (1 - \frac{1}{N} \bigg ) F(t) \bigg \} (1-u)^2,
$$
which leads to the the equilibrium average value
%Kimura and Crow find that the equilibrium condition satisfied by the quantity $F$ is
\begin{equation}\label{kimcrow}
	\< F \> = \frac{1 -  2 \, u }{ 2 \, N \, u - 2 \, u +1 } \approx \frac{1}{2 \, N \, u+1}.
\end{equation}
%(in this paper we choose to disregard the differentiation between actual and effective population size, for simplicity's
%sake).

%Since our population is partitioned into two phenotypes, each of the two phenotypes is characterised by an inbreeding
%coefficient: let us label them as $F_1$ and $F_2$.
%
%[HERE]

The ``infinite alleles" assumption, according to which each new mutation produces a genotype not contained in the population, is
equivalent to assuming that the length $L$ of our genotype is very large, and we shall be making the same assumption
in order to generalise (\ref{kimcrow}).

The line of reasoning used to obtain (\ref{kimcrow}) can be extended to our situation, where a haploid population 
subdivided into two phenotypes $P_1$ and $P_2$ changes by single death-and-birth events, rather than at discrete 
generations: since this setting is less symmetric the details are more articulated, though the idea 
behind the calculation remains the same.

%If we define $F_1$ to be the inbreeding coefficient for $P_1$, we have that the change in $F_1$ from time
%$t$ to time $t+1$ obeys the following recursive relation:

In our model the change in $F_1$ at time $t$ depends 1) on whether a death has taken place, 2) on the phenotype of the organism
that died, and 3) on the phenotype of the newborn (taking into account the possibility that a mutation might have taken place).

\begin{figure}[t]
	\centering
	\includegraphics[width=14.5 cm]{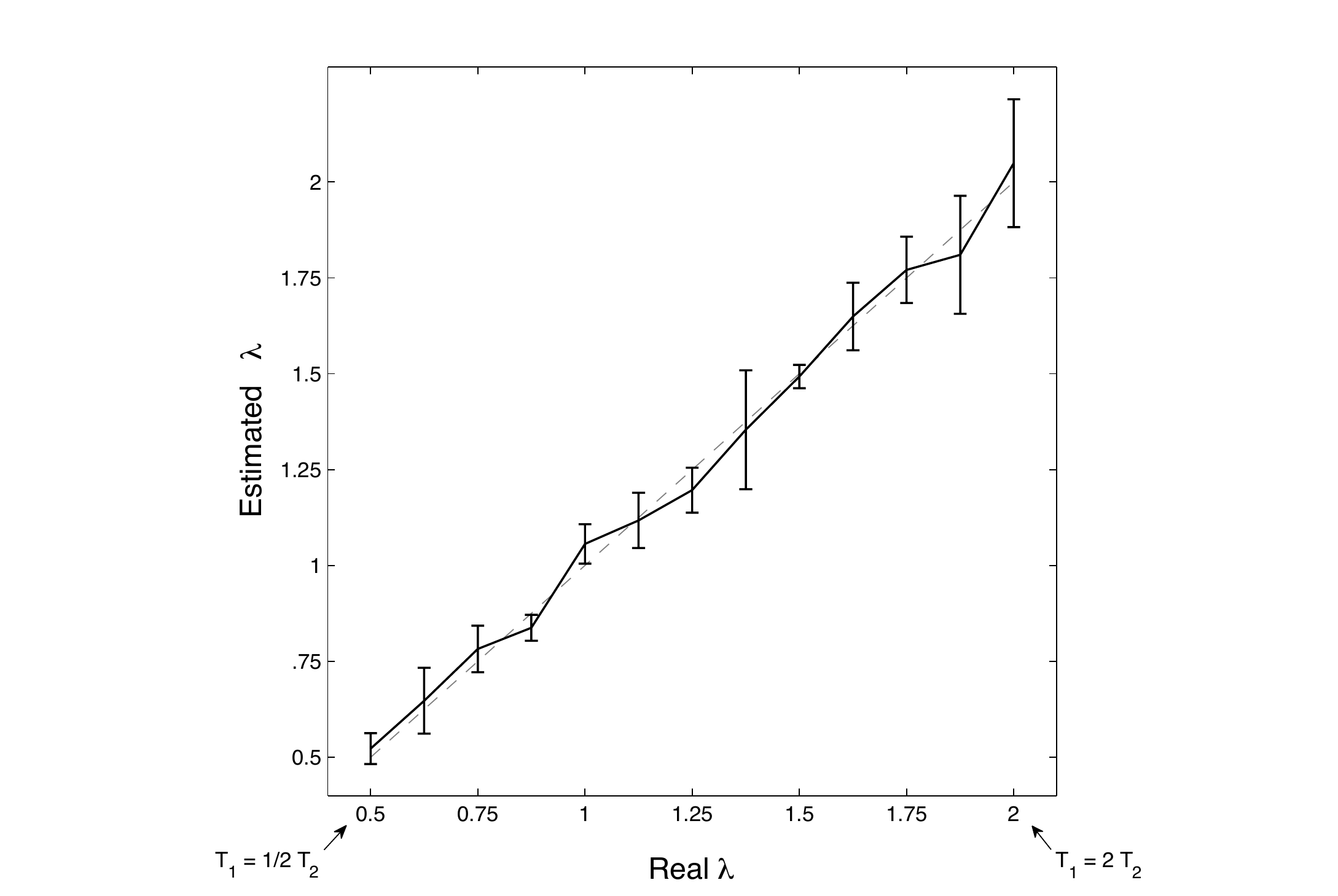}%
	\caption{\it Comparison of the actual value of $\l=T_1/T_2$ and the value estimated from Eqn. (\ref{lambda}), for simulations
		     using values ranging from $\l=.5$ to $\l=2$. The moments needed for Eqn. (\ref{lambda}) are estimated from 10000
		     process realisations for each value of $\l$; the other parameters are $s=-5$, $u=0.007$, $N=1000$
		     and $L=40$.}
	\label{estimapic}
\end{figure}

To reduce the complexity of the calculation we make the assumption that a newborn, before mutation, shares the
same phenotype of the organism which has died. This would only strictly hold if the relative sizes for phenotypes $P_1$ 
and $P_2$ stayed a fixed throughout the process:
in Figure \ref{estimapic} we show, however, the result of a simulation that suggests that our assumption leads to a formula which 
gives a rather precise approximation, and this is sufficient to support the paper's claim that it is theoretically feasible
to extend the population-genetical characterisation of biological function. 

To compute the change in $F_1$ we therefore need to consider three cases for the type of event taking place at time $t$:

\vskip .15 in

\begin{itemize}
	\item[{\bf \emph{A}: }] an organism of type $P_1$ dies, is replaced by a newborn of the same type, and the newborn might mutate,
				  either at the phenotypically-linked site, or at one of the neutral sites,
	\item[{\bf \emph{B}: }] an organism of type $P_2$ dies, is replaced by a newborn of the same type, and the newborn might mutate,
			       	either at the phenotypically-linked site, or at one of the neutral sites,
	\item[{\bf \emph{C}: }] no death takes place, so no replacement happens.
\end{itemize}

\vskip .15 in

According to our process, and taking into account our simplifying assumption -that sets the phenotype of a dead organism
equal to that of the subsequent newborn- the probabilities of events $A$, $B$ and $C$ are:
$$
P(A) = \frac{x}{T_1}, \quad P(B) = \frac{1-x}{T_2},  \quad P(C) =1 -  \frac{x}{T_1} - \frac{1-x}{T_2}. 
$$

Keeping in mind that the quantity $F_1$ is defined as the probability that two organisms chosen at random 
from the population and which have phenotype $P_1$ also share the same genotype, we now need to find 
how $F_1$ changes in each of the three cases $A$, $B$ and $C$.

\vskip .35 in

\centerline{------------------------}

\vskip .15 in

\noindent {\bf Preliminary calculation of sampling probabilities:}

The probability $F_1(t+1)$ is associated with a couple of organisms drawn at random from the population,
and therefore it depends on whether one of the two sampled organisms happens to be the one which was born 
during the last step.
In particular, we need to distinguish the following three sampling events:
\begin{eqnarray*}
	S_1&:&  \textrm{the newborn organism is chosen in the sampling, but its parent is not},
\\
	S_2&:& \textrm{the sampled couple consists of the newborn and its parent},
\\
	S_3&:& \textrm{the newborn is not sampled}.
\end{eqnarray*}
Assuming that the organisms are sampled from the population without reinsertion, the probabilities of events $S_1$,
$S_2$ and $S_3$ are as follows:
\begin{eqnarray*}
	\P(S_1) &=&  \frac{1}{X} + \frac{1}{X-1} -  \frac{3}{X(X-1)},
\\
	\P(S_2) &=& \frac{2}{X(X-1)},
\\
	\P(S_3) &=& 1 - \frac{1}{X} - \frac{1}{X-1} + \frac{1}{X(X-1)},
\end{eqnarray*}
where, like before, $X$ is the total number of individuals with phenotype $P_1$.

Using these probabilities we can now find $F_1 (t+1)$ in each of the three cases $A$, $B$ and $C$

\vskip .15 in

\noindent {\bf Case \emph{A}:}

\noindent To see how $F_1$ changes after an event of type $A$ we need to know two things:
\begin{itemize}
	\item[-] whether the newborn mutated (and whether the mutation happened at the site linked to the change in
			phenotype),
	\item[-] whether one of the two organisms which are selected at random to compute the probability $F_1$ is the newborn 
		     (and whether the other happens to be its parent organism).
\end{itemize}

If a mutation happens at the phenotypically-linked site, any two organisms of type $P_1$ sampled at time $t+1$ will have the same 
probability of sharing
the same genotype, as they did at time $t$: since the probability of mutation at any site is $u$, this will contribute
$$
u \, F_1(t)
$$
to the average value of $F_1$ at time $t+1$, conditional to event $A$.

If a mutation happens at a neutral site, the probability of the newborn having the same genotype as any of the other
organisms is zero (this is a consequence of assuming that $L$ is large enough for each neutral mutation to produce
a totally new genotype). Since the probability of a neutral mutation is $(L-1) \, u$, we have that in this case the contribution
is
$$
0 \cdot (L-1) \, u  \, \big ( \, \P(S_1) + \P(S_2) \, \big ) +  F_1(t) \cdot (L-1) \, u \, \P(S_3) = 
F_1(t)  (L-1) \, u \, \P(S_3),
$$
The first term on the left-hand-side corresponds to the event that the newborn is chosen at the sampling (events $S_1$ and $S_2$ above),
whereas the second term, which gives the non-zero contribution, corresponds to the fact that the probability $F_1$ remains 
unchanged as long as none of the chosen organisms is the newborn (event $S_3$).

Finally, for the case in which no mutation happens at any site, which has probability $(1-L \, u)$, we get the following 
contribution:
$$
(1-L \, u) \big \{  F_1(t) \cdot \big ( \P(S_1) + \P(S_3) \big ) + 1 \cdot \P(S_2)  \big \},
$$
where the second term corresponds to the fact that, as long as no mutation takes place, the probability that the newborn 
shares its genotype with its parent is equal to $1$.

Therefore, if we denote the value of $F_1(t+1)$ conditional to event $A$ by $F_1(t+1|A)$, summing all three contributions
we get
$$
F_1(t+1|A) = u \, F_1(t)
		 + F_1(t)  (L-1) \, u \, \P(S_3)
		+ (1-L \, u) \big \{  F_1(t) \cdot \big ( \P(S_1) + \P(S_3) \big ) + \P(S_2)  \big \}.
$$

\vskip .25 in

\noindent {\bf Case \emph{B}:}

\noindent In this case we have a newborn of type $P_2$. Like for case $A$, we use the term $F_1(t+1|B)$  to the denote the new 
value of $F_1$ conditional to $B$, and we have that
$$
F_1(t+1|B) =  \bigg ( 1 - u \, \big ( \P(S_1) + \P(S_2)  \big ) \bigg ) \, F_1(t).
$$
It's straightforward to see why: if the newborn is of type $P_2$, the inbreeding coefficient remains unchanged unless the newborn
mutates in the phenotypically-linked site and is subsequently chosen in the sampling: the probability of this event is
$u \cdot \big ( \P(S_1) + \P(S_2)  \big ) $.

\vskip .25 in

\noindent {\bf Case \emph{C}:}

In this case nothing happens, so we have that
$$
F_1(t+1|C) = F_1(t).
$$

\vskip .15 in

\centerline{------------------------}

\vskip .35 in

We can now write the value of $F_1(t+1)$ as the sum of the conditional contributions multiplied by their respective
probabilities:
$$
F_1(t+1) = F_1(t+1|A) \, P(A) + F_1(t+1|B) \, P(B) + F_1(t+1|C) \, P(C),
$$
and we can use symmetry to obtain an analogous relation for $F_2$.

In the limit of large $N$ these two relations simplify substantially, so that up to order $1/N^2$ we get
\begin{eqnarray*}
	F_1(t+1) - F_1(t) &=&
				 \frac{2}{N^2 \, T_1} \, \Bigg \{ \frac{1}{x} - \frac{1}{x} \, F_1(t) \Big ( 1 + \theta  \big (  \l (1-x)+x \, (L - 1) \big ) \Big )  \Bigg \},
\\
	F_2(t+1) - F_2(t) &=& 
				\frac{2}{N^2 \, T_2} \, \Bigg \{ \frac{1}{1-x} - \frac{1}{1-x} F_2(t) \bigg ( 1 + \theta \Big ( \; \frac{ 1 }{ \l } \, x+ ( 1 - x ) (L - 1) \Big ) \bigg )  \Bigg \},
\end{eqnarray*}
where the terms $x$ and $(1-x)$ at the denominator arise from the sampling probabilities $\P (S_1)$, $\P (S_2)$, $\P (S_3)$.
%are a consequence of the subpopulation size for the two phenotype being $N_1 = x \, N$ and $N_2 = (1-x) \, N$.

Therefore, denoting by $\< \cdot \>$ the average with respect of all realisations of our process, we get the following relations linking 
the moments of the observable quantities
to the model parameters:
\begin{eqnarray*}
	\big \< F_1 / x \>  + \theta \Big (  \l (\< F_1 / x \> - \< F_1 \> )+ (L - 1) \< F_1 \> \Big ) &=& \< 1/x \>,
\\
	 \< F_2 / (1-x) \> + \theta \Big ( \, \frac{ 1 }{ \l } ( \< F_2 / (1-x) \>  -  \< F_2 \> )  + (L-1) \< F_2 \>  \Big ) &=&  \< 1 / (1-x)  \>.
\end{eqnarray*}

Though the notation is somewhat cumbersome, it is easy to see that these two equations offer a relation between the model 
parameters $\theta$ and $\l$ and the averages of the six random quantities
$$
 F_1, F_2, \frac{1}{x}, \frac{1}{1-x}, \frac{F_1}{x}, \frac{F_2}{1-x},
$$
all six of which are in principle statistically observable.

Since this paper focuses on the parameter $ \l = T_1 / T_2 $, in virtue of its putative relevance in terms of function, we 
proceed by solving both equations for $\theta$, and equating them in order to find a relation for $\l$.

In order to express the mentioned relation in a more compact form we define the following auxiliary quantities
$$
R = \frac{ \< F_2 \> }{ \< F_1 \> } \cdot \frac{  \< 1/x \> - \< F_1/x \> }{ \< 1/(1-x) \> - \< F_2/(1-x) \> },
$$
$$
Q_1 = \frac{ \<F_1 / x \> }{ \< F_1 \> } - 1, \qquad Q_2 = \frac{ \<F_2 / x \> }{ \< F_2 \> } - 1.
$$

In terms of these quantities, the equation for $\l$ takes the following form
$$
R = \l \frac{ \l Q_1 + L-1 }{  Q_2 + \l ( L-1 ) },
$$
and this relation leads to a quadratic equation that only admits one non-negative solution:
\begin{equation}\label{lambda}
	\l = \frac{ 1 }{ 2 Q_1 } \bigg \{ (R-1)(L-1) + \sqrt{ (R-1)^2(L-1)^2 + 4 R Q_1 Q_2  } \bigg \}.
\end{equation}

Figure \ref{estimapic} shows the result of using formula (\ref{lambda}) to estimate $\l$, for a series
series of simulations where the real value of $\l$ ranges from $.5$ (i.e. $T_1 = 1/2 \, T_2$) to $2$
(i.e. $T_1 = 2 \, T_2$): we see that the average values of such estimations are well aligned with
the actual values.

The magnitude of the standard deviation for our estimations, on the other hand, is considerable, especially
in view of the fact the 10000 realisations of the process were used to estimate each value of $\l$: it is clear
that a substantial increase of efficiency will be needed to make the theory relevant to actual empirical
phenomena.

This practical consideration should not obfuscate, however, the fact that equation 
(\ref{lambda}) provides a relation between population statistics given by
$x$, $F_1$ and $F_2$, and parameter $\l=~T_1/T_2$, which contains functional information
related to a gene's effect on an organism's ability to survive, rather than on its
 reproductive fitness.

\section{Outlook}

We have shown that differentiating the lifetimes of two phenotypes independently from their 
fertility leads to a qualitative change in the equilibrium state of a population: since 
survival and reproduction are quite distinct macro-functions performed by any living organism,
this contributes to extend the population-genetical characterisation of biological function.

We have furthermore shown that, by using information provided by neutral variation, 
the lifetime ratio $\l$ can be expressed explicitly in terms of statistically observable 
quantities, and independently of all other parameters. 
%This both sets up the model as one which is, in principle, 
This both gives some support to the possible empirical relevance of the proposed modelling 
approach, and 
suggests observable quantities that can be useful in characterising the stochastic equilibrium 
of a population in terms of the functional features of the individuals which comprise it.

It needs to be stressed, however, that the statistical resolution needed to estimate $\l$ efficiently
following this method seems to go beyond what could be achieved empirically: in order to obtain Figure 
\ref{estimapic}, 10000 realisations of the system were needed for each parameter value, and for each  
value 5000 generations were needed for the population to relax to its stochastic equilibrium.

This study aims to be a proof of principle, and should only be considered a Òworst case scenario,Ó which 
nevertheless shows that inferring functional details from population genetical considerations is a definite 
theoretical possibility.
It is left for a future work to assess its practical feasibility by improving the estimation efficiency, 
possibly while considering dynamical statistics explicitly:
it is useful to remember, however, that the dynamics of no system has ever been understood without a sufficient 
grasp of how relevant forces balance one another to allow observation.

%[OBSERVATIONS ON MAXIMUM LIKELIHOOD]

\vskip .2 in

%\noindent {\bf Acknowledgments}  \
%I'd like to thank Prof. Henrik Jeldtoft Jensen for drafting an ecological model which partially inspired the one 
%presented here, and Prof. Rick Durrett, as well as two anonymous reviewers, for detailed comments on 
%a preliminary manuscript. 
%Thanks are also due to Prof. Pierluigi Contucci for pointing me towards this type of modelling,
%% to Prof. Ping Ao and Dr. Shuyun Jiao for inspirational conversations that lead to a change in my analytical approach, 
%and 
%to Ricardo and Federico Gallo for the scientific influence that led to this paper. The research was supported 
%by a Marie Curie Intra European Fellowship within the 7th European Community Framework Programme.

\noindent {\bf Acknowledgments}  \
An acknowledgment is due to Henrik Jeldtoft Jensen for drafting an ecological model which partially inspired 
the one presented here. The research was supported 
by a Marie Curie Intra European Fellowship within the 7th European Community Framework Programme.

\end{document}